# Critical Aspects of Modern Open Source Software Technology to Support Emerging Demands

Umer Farooq
Department of Software Engineering, FUIEMS, Rawalpindi Pakistan

M. Aqeel Iqbal
Department of Software Engineering, FUIEMS, Rawalpindi Pakistan

Usman Shabbir
Department of Software Engineering, FUIEMS, Rawalpindi Pakistan

Sohail Nazir
Department of Software Engineering, FUIEMS, Rawalpindi Pakistan

**ABSTRACT**
Software has gained immense importance in our everyday life and is handling each and every aspect of today's technological world. The idea of software at initial phase was implemented by a very precise minority of individual and now it's everywhere whether one's personal life or an organization .Financially strong organization and people who can purchase this bounty of technological era can fulfill their desires efficiently. For sure it's not a generalized case that one is financially strong and can easily afford the desired software. There are numerous users who cannot do so. Open source software has a way out for these users it provides them the same facilities and functionalities as in their equivalent software irrespective of any financial pressure. So the financially constrained personals or organization can make use of open source software for achievement of their desired tasks. In this research paper an analysis of open source software has been presented by providing a brief comparison of Ubuntu as an emerging high quality open source modern operating system with well known Microsoft windows operating system.

## 1. INTRODUCTION
The role of software in today's technical world is evitable. Technology has taken over even the very basic tasks that where to be done manually by humans to control the behavior and to make the best use of technical tools software was required. As with the passage of time with the transformation of hardware the concurrent development in software domain was also required but due to critical role of software the transformation of technological development its cost was boosted. These financial restrictions narrowed the domain of technological progress. As a net effect the pace of development was immensely reduced any it became the royalty restricted to the one's having no financial restriction. Such scenario need to thought that the financially suppressed user's shell also be making use of software enhance the performance. [1]

The procurement of the software was easy for the organizations that had sufficient financial budgeting. As the heavy financial load bearing was not generally applicable so the low budget organizations going for performance software had to module all that in their financial domains [2]. So the thought arose among the stake holders of the software industry that the software usage shell not be restricting the users on ground of financial limitations. Open source software generally refers to availability of the source code with all the rights normally reserved by owners [3]. The software license shell allowing modification in the code, in order to improve the software performance and to keep it in touch with the constantly changing software requirements how the software is set to be open source provided that it's license holds the criteria's that generally adopted the Debian free software guides, that examines whether certain software licenses is a free software license [2]. It's the free software license which gives the user rights of modification of software along with its distribution. Free software refers to the freedom in the shape of permission to further distribution of existing copyright work. There are certain conditions that are required to be fulfilled by a license of a particular software so shall it could be consider an open source one.

### 1.1 Free redistribution
There shall be no restriction imposed on any individual customer so that the further distribution of a particular code either is original or the modified form, done individually or as a colorable effort is restricted with no royalty claim for any such redistribution. [3][4]

### 1.2 Source code
The deliverable must be containing source code, with redistribution permission. In both the source and it's complied form. If product launched does not posses the source code, there shall be an easy and approachable mean defined for accessing the source code that could be such as internet website without financial claim. Simultaneously the source code shall be in a form easy for user to modify. Making the source code unclear or hard to understand and the semi functional code is restricted.

### 1.3 Derive work
The license shall grant user permission to allow modification and to make use of modified code as component of original software's license terms and conditions.

### 1.4 Author's source code integrity
Further in hand distribution of software could be restricted for further distribution in modified form particularly if the permission to distribution of patch files for modifying code at the time of building. The license shall be clearly permitting the distribution of modified code. In case of derived work a license may be binding the modified to further distribute work to be released as different version number or name.

### 1.5 No discrimination against person or group
No restriction could be imposed on particular person or a group on any base so that software could be utilized by them.

### 1.6 Discrimination avoidance against endeavors
The license should not be restricting the use of software to be utilized for a particular field.

### 1.7 Distribution of license
The right authorize to the user in the original software shall be applied to all others to whom software is further distributed.

### 1.8 License shall not be specific to a product
The rights allotted with the particular software shall be domain till the software is being used for a particular product or field. In case any such program is obtained from distributor and utilized for some other purpose. Then all those to whom software shall be further distributed shall have the same tights as hold in original license. [3]

### 1.9 Other software must not be restricted by the License
License shall be strictly not being making binding on other software that is distributed concurrently with it to be an open source one. [4]

### 1.10 License must be neutral from technological perspective
No component of the license shall make the software difficult to be used for a particular technology or interface styling. Apart from the mentioned criteria in order to get license officially recognized there are other condition that software shall be fulfilling. These include unique name, legally satisfying the open source development criteria and eventually getting an approval from open source initiative. Examples of open source license include BSD (Berkeley Software Distribution), GPL (GNU Public License), LGPL (Lesser General Public License), MIT (Massachusetts Institute of Technology) and MPL (Mozilla Public License). Table 1 gives a brief comparison of these. [5]

|  | Notice of modification | Redistribution of the modified work | Linking to closed source code |
|---|---|---|---|
| BSD | Yes | Yes | Yes |
| GPL | Yes | Only under GLP or LGPL | No |
| LGPL | Yes | Only under GLP or LGPL | Yes |
| MIT | Yes | Yes | Yes |
| MLP | Yes | Only under MLP | Yes |

**Table 1: A comparison of open source software license**

## 2. DEVELOPMENT PHILOSOPHY
Traditionally software was developed the way cathedrals used to be constructed in the ancient era. For doing so skilled artisan used to plan their tasks individually and was later on implemented by an individual effort. As the task was done a few modification required were done. Software construction was following such mode of development. Individual or a group of programmers used to work isolate with precise and accurate planning and management once the task was accomplished the product was released to the world with subsequently maintainace as required was done.

Bazaar model is considered to be platform for the philosophical grounds of open source software. by bazaar it is inferred that initially traders comes established their shop and ware houses and then initial business .Subsequently more traders come and established their structures as a net effect the bazaar grows .Initialized as a very basic structure, the structure is modified as the requirement start to furnish .Similarly in open source software developer initially release a prototype and then build the final product based on demand of the end users. other users could also play their role in the software development on the existing state of code .After certain span an entire desired product is achieved and later on evolves periodically .The bazaar model is being followed by a major products such as Mozilla Linux etc. following patterns that a software development shall full fill in order that it could be considered a bazaar model platform product which are as under. [5]

### 2.1 Treatment of a user as a Co developer
Users are tackled as co developers they are given access to full source code a long with the software and are welcomed to provide their feedback. As a result the increase in co developer indeed accelerates the rate of software evolution, as due to more co-developers bugs, errors etc. are easily identified and encountered. So users who posses programming skill can even perform better they identify the flaw and suggest the solution along with it and can also plays role as an additional tester.

### 2.2 Early release
The initial version of software shall be immediately released, so that more co-developers could be found.

## 2.3 Frequent integration
The modification shall be introduced immediately as integrating few modification is easy than to do all integration all together. [6]

## 2.4 Several versions
There shall be two categories of released software concurrently one with a high functionality and bugs for the users who intend to make use of software immediately and willing to face risk and the other one with a limited functionality and so with a fewer bugs.

## 2.5 High modularization
The component or the module based software development shall be adopted to allow parallel development.

## 2.6 Dynamic decision making
There shall be a decision making setup, that could be formal or informal may be official or un-official to make decision the changing user demand, stake holder alter opinion and other issues concerned the software development [2][6]. The bazaar model inherits some advantages which are as under.

## 2.7 Effort reduction
As the open source software is released early with early rights granted to user. The task being done on a particular module is being done concurrently by many individuals that yield perfection, efficiency and reliability involving thousands of developers round the globe.

## 2.8 Utilization of other's effort
As the source code is available so if a developer requires building software with similar functionality he makes use of existing source code. In scenarios when a code cannot be integrated as it is, even then the time is saved that was meant to be consumed while designing other similar software. [3][5]

## 2.9 Enhanced quality with control
As the bazaar model's involves a large number of people as coworker so the bugs are identified and managed more efficiently an easily. Some users are granted access not only pinpoint it but in some cases may even suggest a remedy for it. Such users also play there role in enhancement of quality and control of software. [7]

## 2.10 Decrease in maintainace expenditures
Maintainace is a process that evolves through out the life cycle of software which consume much of financial resources in case of the open source software based on bazaar model potential users, a variety of maintainace solutions are available so the best suited one is adopted. This availability leads to competition that decreases the cost. [5][7]

## 3. PROS AND CONS
Aside from the financial privileges attained there are numerous benefits of open source software. Few of them are explained as under.

## 3.1 Reliability
Reliability can be used for multiple purposes as it's a loose term. In software perspective by reliability we refer the absence of defects that could cause a software failure. This failure is often regarded as a bug. As by default the users or the co developers are in large number when it comes to open source software due to which the bugs are identified and dealt with amazingly with efficiency and speed. [5]

## 3.2 Stability
As once functional the software requirements change that require the software to adopt the desired changes. As in case of open source software the source code is provided along with the software it makes the evolution of the software easier for the user. [7]

## 3.3 Auditability
This is the benefit of open source software known by minority. Imagine if one goes for a closed source software user have to trust the vendor .when certain specification and features are claimed. On contrary by passing the source code along with the software .the basis for claim is provided. This is for sure that no audit could be done without the source code that is done specifically open source software product.

## 3.4 Financial issues
Generally the open source software is for free. Free could be expressed in two meaning one free software and the other free in terms of source code .there are some exemptions to it but main open source software are satisfy the both meanings of free. Glittering of the open source software tempts everyone toward it but in order to utilize this and for a successful software there are certain areas that are require to be simultaneously satisfied which are difficult to be applied generally.

## 3.5 Software skills
A software user shall be having sufficient knowledge of the software platform and coding in order to utilize the code distribution along with the software. Else the distributed code is just an accessory attached which is claimed to be beneficially but could not be utilized. For which one has to have sufficient skills in the software coding implementation and knowledge areas which is for sure not a generalized scenario.

## 3.6 Process clarity
The open source software is at a single instance being built on different platforms, each having respective goals and perspective further more the same goals and objectives are implemented by various developing communities this parallel development each at different stage and being build by different creation methodologies makes the software development process unclear.

## 3.7 Dead and user
As claimed the user community works as co developer that is assume to be among the most appealing characteristics of open source software. What if you launch software and could not attract the users and eventually lead your software to a failure.

Such scenario generally occur due to non compatibility of the objective ,development models and goals of actual developer and the co users that could be proved fatal for software lifecycle.

### 3.8 Balance between user and developer.
While constructing software intended to be an open source one, the developer has to focus on attracting the co developers for success of open source software. In doing so, the developer had to add certain features that tempt the coworkers so that he works for the development of software. Such features may lie beyond the scope or motive of original software. So in getting a successful open source software requires a balance between the developer and user acting as co developers which some time may as tray software development path from original developer's goal. [8]

### 3.9 Niche software
In an open source software development the software development procedure is greatly influenced by the likeness and dislike of a user playing role of co developer. So is the further development of it is effected and as the co developer might develop software based on personal areas of interest that cannot be compatible with the original developers goal. [9]

### 3.10 Regulation blame
In the development of open source software there are multiple versions of software being developed which could be further characterized on the motive, construction model adopted etc. In such scenario as there are multiple developments being done that too on multiple platforms? In such scenario the issue of regulation blame may arose for which no one could sorted out to be blamed.

### 3.11 After sale services
This is most vital phase of software development and it's the phase when the software shows it worth. In case of open source software only a few selected products gave the guarantee the proper functionality of software and the majority puts it on luck that will it perform or not.

### 4. MARKETING
When it's come to marketing it's the most important factor that leads to software success in perspective of market share and user community using software. In contrast the open source software with the propriety software's one concludes the features are almost equivalent rather in certain domains the open source software gains an edge. Analyzing windows and Ubuntu .Is windows a superior product, the answer is a clear cut no so what's that Ubuntu is known by a minority and the windows by a clear and domain and majority. The answer is marketing following are the considerations that shall be adopted for successful marketing.

### 4.1 Span of launching software
The companies who initiate a new dimension to a market are always at the lead, having an edge over other competitors to be the first to implement certain software. Such scenario can boost success if initiated with a large installed base channel and regulating their software development in right directions. [11]

### 4.2 Advertising
All efforts and efficiencies of software aside though have a vital role in getting customers reorganizations. But the backbone is creation of awareness of existence among the potential user communities that could be achieved by elicitation techniques as the advertising is a main task. For sure requiring a hefty financial backup as the open source software product could be restricted due to financial constrains as majority are not commercial products , However there are certain advertising modes that may need no or even if require may need low investment of capital. The mediums could be internet, official websites, magazines networking, seminars, workshops etc.

### 4.3 Trade mark
Brand or the trade mark is some thing that distinguishes a product manufactures from the other one. It's the identity that reflect ones pre launched products and achievements that plays a vital role in marketing or the success of software as it's your trademark that distingue and is the reference that would we referred while recommending or advertising product. [10]

### 4.4 Scale
The user of the software are generally unaware of the technical tools and platforms on which the software is constructed so instead of technical description of software job it shall be explained in terms of business value of the software. [11] Such efforts makes the customer more aware of what the software is all about and how its usage could help them in efficient manage of task.

### 4.5 Distribution
Distribution of software shall be done in a manner that it reduces the complexity of user in interpreting the software usage and application. The common case is of Linux and Linux under Redhat supervision. Redhat emphasized on user easiness that led to success of Redhat Linux with respect to other version of Linux.

### 4.6 Way out
The way out also referred as exit strategy is the planning of tasks that shall be implemented on achievement of goals. Steps taking generally are selling of software or redefining of goals to remain in touch with the changing software requirement.

### 5. OPEN SOURCE SOFTWARE DEVELOPMENT
Software development governance is an very important factor that manages all the concerns during the entire software lifecycle . if goverened efficiently a software could be a success full one on contaray if ill managed may result into a disaster with respect to development goal. The software governance involves some mjor concerns which include management at organisational level , risk management e.t.c. other concerns include the flexibility in the software development , measure taken inorder to minimize risks and to satisft the furnishing compliance needs that could be internal or external [11]. As mentioned earlier open source software development is a collaborated process in which user act as co-developer. Figure 1 shows a development cycle of open source software. However in order to know an open source software development we must be having knowledge of following terms.

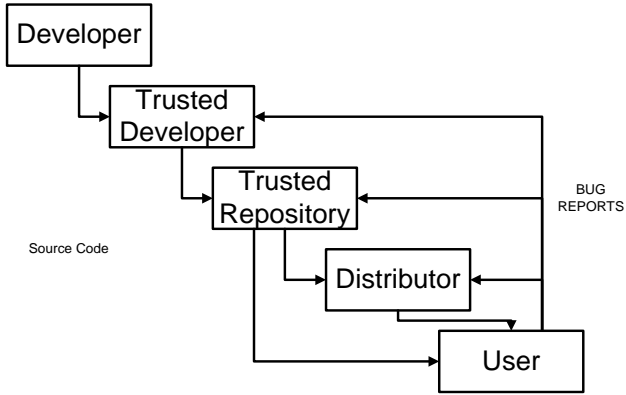

**Fig 1 Shows a development of open source software along the input from user who act as a developer.**

## 5.1 Trusted repository
A trusted repository refer to some web location from where the users who intend to get particular open source software can get the official version or the official release of the software along with the other components such as source code, documentation related to software etc.[7] A user who wishes to get open source software can get it directly from the software trusted repository or may contact the distributors who get it from the trusted repository then modify it as per user demand that may be integration demand, testing special configurations etc.

## 5.2 Trusted developers
Tested developers include the developers who initiate an open source software project and create repository. They can further figure out the trusted developers among the users of software on account of the expertise and contribution to the software development.

## 5.3 User
Users are the potential customers of the software. They can send bug reports or any other anomalies in the software to the trusted repository or the trusted developers. The user has the authority to modify local copy of software either personally or by a proprietary programmer use.

After attaining the software and the source code from the trusted repository or from a distributor, open source software user has authority to modify a program and could even post their modified versions to the internet but in order to get their modified open source software to be published on the platform of trusted repository, The software is tested and analyzed, if observed that the modified code is of worth publish it as official version.

Imagine open source software released having certain bugs. As it would be released it would attract users towards it out of them some may play role of co developers based on inputs from co developers to the trusted repository of the particulars open source software. The bugs would be identified and corrected. As the improvements done by others would be tempting more users towards the open source software as by nature people are habitual of making user others effort and then even if they plug-in their feedback an avalanche like progress would be observed yielding efficient results under prospective of time, quality and financial load as would be shared among a huge amount of users.

The most time consuming part of the software development is the state of vendor lock in, which could yield fatal results when it comes to quality software. As the open source software development cycle involves a lot of co developer. So, if such a condition arises there are many eyes and heads focused on software, which for sure eliminate the chance of vendor lock in. This feature also reduces the cost factor, as a variety of solutions that to from multiple sources are available, that eliminate the vendor lock in which is the major cause of rise in cost substantially. [3]Asides a user would like the changes to be done by others and the only way it could be done is by pointing out the bugs in the software to the software repository, which would take further steps for removal of those anomalies this.

Apart from providing the solution to the anomalies, as the same would be done by other users as a net result user would be enable to make use of all improvements instead of their own. By repetition of such scenario the software quality polishes more and more and eventually more and more user community is attracted which further increases the potential users that in turns plays vital role in a quality software development. Generally there exists a misconception regarding the developers of the open source software that's what if the developers are inexperienced. The sure shot answer is the particular open source software would crash, but the reality of statistics is yielding astonish result that is the average age of open source developer is thirty year with in average experience of eleven point eight years. [3]

## 6. A BRIEF SURVEY OF UBUNTU OPERATING SYSTEM
Ubuntu is open source software from the Linux family of operating system. It is a community based operating system having features that make it perfect for all platforms to be used as an operating system, whatever it may be a desktop, server or a laptop. Although free but with the superb features of periodic security updates after eighteen months and a new release after six months and all that for free. Having extra ordinary feature of live cd running and ready to use environment the moment installed makes Ubuntu compatible with any operating system.

### 6.1 Governance
The governance of Ubuntu operating system is conducted in the following levels:-

#### 6.1.1 Local teams
Governed in an efficient manner with the collaboration of local teams whose task is working with local level Ubuntu user groups and with the government in order to make people aware of this operating system bounty.

#### 6.1.2   Technical board
Technical board responsibility includes package selection, kernel, and library versions extra. This technical board meets regularly after two weeks.

#### 6.1.3 Ubuntu community council
Management of social structures and community processes of Ubuntu are done by the Ubuntu's community council. The other primary task of Ubuntu community council is the dispute resolution between the local teams. The Ubuntu community council meets after a period of two weeks.

### 6.2 Components
On account of ability of support by Ubuntu or on the basis of the software goal with relevance to open source philosophy the components are characterized as under.

#### 6.2.1 Main component
All software's that are free could be redistributed irrespective of any royalty claim and are completely supported by Ubuntu official platform lie in main component. Majority of main components are pre installed by default in it. The software under this category is considered to be the most reliable ones.

#### 6.2.2 Restricted component
This category includes the software that are most often utilized and supported by Ubuntu. They may or may not fall under the category of free license.

#### 6.2.3 Universe component
This category includes the software from Linux world and open source software. It includes almost all open source software and other software that reside outside the category of open source software. The Ubuntu is not providing guarantee, security fixes and support.

#### 6.2.4 Multi-verse components
All software that is not free falls under the domain of multiverse category and they don't satisfy the Ubuntu's main component license policy.

### 6.3 Application and uses
Applications of Ubuntu operating system involves all the features that are expected to be in good operating system including access thousand of open source software including games, education, sound, streaming videos, email and chat office applications, sync and memory sharing etc.

## 7. Benchmarking Ubuntu 9.04 and windows vista
Linux operating system family especially Ubuntu have enhanced their quality and as a result attracted a lot of user community. On account of features offered Ubuntu has appeared as an alternate for Microsoft windows. But despite equivalent features Ubuntu lags for behind rather out of consideration with respect to Microsoft windows share in operating system usage worldwide.

As the survey reflects that ninety percent of the operating system shared worldwide is dominated by Microsoft windows and the remaining ten percent is shared among various other operating systems including Ubuntu. With the passage of time the quality of Ubuntu has flourished up to a great extend and eventually reach a stage where it has become alternate for Microsoft windows. Under the light of certain aspects Ubuntu has taken an evident edge over Microsoft windows but still lacks behind rather for behind in attaining user community as windows had.

### 7.1 By default installation on computers
A lot of marketing emphasis done by Microsoft has created a scenario under influence of which many of the launched computers have by default Microsoft windows operating system installed on. As a result majority of computer user are aware of Microsoft windows as the only operating system to exist. We grow in the environment surrounded by Microsoft operating system all around.

### 7.2 Knowledge of operating system
We are in use with the environment having Microsoft operating systems are the only one majority of aware of. As a general influence exist that the free counter parts of the proprietary operating system are of low quality. Windows vista despite being an outclass operating system its alternate Ubuntu 9.04 is also and equivalent to it but on account of knowledge of operating system could not get as many users as vista holds.. A brief comparison of Ubuntu and windows vista is in table 2.

### 7.3 Marketing
Due to non centralization of Ubuntu the marketing is done individually by different versions and user communities which are poor marketing strategies. Alternately Microsoft has a centralized organization having a dedicated marketing department.

### 7.4 Understanding of operating system
As the majority of users are concerned with the working of their task by the operating system their unconcerned with the technology of the operating system and the not concerned to search for an alternative windows.

| Feature | Ubuntu | Vista |
|---|---|---|
| Versions/ editions | Desktop Server | Home Basic Premium Business ultimate |
| Price | Free | Not Free |
| Built in security<br>    i.    antivirus<br>    ii.   firewall | <br>Yes<br>Yes | <br>Yes<br>Yes |
| Updating | Yes | Yes |
| Relative performance on common hardware | High | Low |

**Table 2 shows a comparison of Ubuntu 9.04 and widow's vista**

### 7.5 Improvement

Claiming marketing to be the only reason for success is not fair, as the marketing could boost sale of a standard and a quality product only. Microsoft has enhanced the quality of their operating system up to a great extent. As a clear evident example is the evolution of windows from Windows 95 to windows 7, for sure it reflects the enhancement of quality and performance. . Table 3 shows the system requirement contrast between Ubuntu and windows vista.

| Requisite | UBUNTU 9.04 | Windows Vista |
|---|---|---|
| Processor | 300 MHz 32 or 64 bit processor | 1 GHz x86 or x 64 processor |
| Ram | 64 MB | 512 MB for home basic 1 GB for Premium, Business, Ultimate |
| Disk Space | 2 GB | 20 Gb for home basic and 40 GB for Premium, Business, Ultimate |

**Table 3: shows the minimum requirement for the system requirements of Ubuntu and Microsoft Vista.**

### 8. CONCLUSION AND FUTURE WORKS

The benefits inherited by open source software have played a crucial role in the improvement of the software that yielded the current scientific computing. Aside the software quality was boosted as a net effect of features of open source software that include enhanced verification, validation and quality. Tough with the merits of open source software variety of de merits had to be encountered, but all this could be addressed in order to minimized or even up to great extent even nullified by the proper management. Both the open source and the propriety software had their respective limitation, but, its all user influenced decision of deciding to go for an open source or propriety software. The decision is strictly dependent on the software user skills and degree to which software performance could be compromised. Predicting about the future it's hard to decide that who's going to dominate the software market. But as one observes the configuration and features of certain open source software products it's concluded that their propriety software counterparts lag far behind. This scenario ignites the thought that if we do not agree that open source software is going to rule the software market we are just making a fool out of ourselves.